\documentclass[10pt]{iopart}

\usepackage[dvips]{graphicx}
\usepackage{iopams}

\renewcommand{\vec}[1]{{\bi{#1}}}

\renewcommand{\Im}{{\rm Im}}

\graphicspath{{figures/}}

\begin{document}

\jl{3} 

\title[Quantum effects \dots Kondo lattice model]{Quantum effects in the
  quasiparticle structure of the ferromagnetic Kondo lattice model}

\author{D. Meyer\footnote{present Address: Department of Mathematics,
    Imperial College, London, United Kingdom}, C. Santos and W. Nolting}
\address{Lehrstuhl Festk{\"o}rpertheorie, Institut f{\"u}r Physik,
  Humboldt-Universit{\"a}t zu Berlin, Invalidenstr.\ 110, 10115 Berlin, Germany}

\date{\today}

\begin{abstract}
A new ``Dynamical Mean-field theory'' based
approach for the \textit{Kondo lattice model} with quantum spins is introduced.
The inspection of exactly solvable limiting
cases and several known approximation methods,
namely the \textit{second-order perturbation theory},
the \textit{self-consistent CPA} 
and finally a \textit{moment-conserving decoupling} of the equations of
motion help in evaluating the new approach.
This
comprehensive investigation gives some
certainty to our results:
Whereas our method is somewhat limited in the
investigation of the $J<0$-model, the results for $J>0$ reveal important
aspects of the physics of the model:
The energetically lowest states are
not completely spin-polarized.
A band splitting, which occurs already for
relatively low interaction strengths, can be related to distinct
elementary excitations, namely magnon emission (absorption) and the
formation of magnetic polarons. We demonstrate the properties of the
ferromagnetic Kondo lattice model in terms of spectral densities and
quasiparticle densities of states.
\end{abstract}



\section{Introduction}
The Kondo model  and its periodic extension, the Kondo lattice model
(KLM), which describe
spin-exchange interaction between a localized spin or a system of
localized spins, respectively, and a band of itinerant electrons,  has been
subject of intense 
theoretical studies in the
past~\cite{Zen51,AH55,Kas56,Kon64,Nag74,Nol79}.  
This model has been applied to a variety of different problems in
solid-state physics using both a ferromagnetic and antiferromagnetic
coupling constant $J$.

The model with $J<0$ is the one originally known as 
\textit{Kondo lattice model} or simply \textit{Kondo model} in its
non-periodic form with a single impurity spin in the system. It was used
by Kondo to explain the unusual temperature behavior of the resistivity
of magnetic impurities in non-magnetic hosts~\cite{Kon64}.
The negative spin-exchange
interaction can be derived from the hybridization of a correlated
``atomic'' level with a conduction band, the situation described by the
Anderson model~\cite{And61,hewson}. In the limit of a low-lying
half-filled atomic level and strong correlations, the Anderson model can
be mapped onto the Kondo model with a negative exchange
constant~\cite{SW66}. The Kondo lattice model is still subject to much
theoretical work, the main objective is the understanding of the
unusual physical behavior found in \textit{Heavy-Fermion}
materials~\cite{hewson}.

A model with identical operator structure in the Hamiltonian, but with
positive exchange constant has been known in the literature for a long
time by many different names 
(double exchange model, $s$-$d$ model, $s$-$f$
model,\dots)~\cite{Zen51,Kas56,Nag74,Nol79}. 
For clarity, we will refer to this model in the following as
\textit{ferromagnetic Kondo lattice model}.
The model with
ferromagnetic exchange has to be understood as an effective one.
The origins of the exchange with $J>0$ are found in the interband
Coulomb correlations~\cite{Kas56}.
This situation is believed to dominate
the physical properties of important systems such as
the magnetic
semiconductors~\cite{Wac79} (EuX; X = O, S, Se, Te), the diluted magnetic
semiconductors~\cite{Kos76} (Cd$_{1-x}$Mn$_x$Te, Hg$_{1-x}$Fe$_x$Se), and
the ``local moment'' metals~\cite{Leg80} (Gd, Dy, Tb).
To these problems, the ferromagnetic KLM was successfully
applied~\cite{NMR96,NRM97,REN98}.
Recently, this variant of the KLM has gained
a lot of 
interest with the discovery of the colossal magnetoresistance (CMR)
materials~\cite{JTMFRC94,Ram97}. In these materials, typically manganese
oxides with
perovskite structure (La$_{1-x}$(Ca,Sr)$_x$MnO$_3$), the double-exchange
model~\cite{Zen51,AH55} has been successfully 
applied to explain the origin of ferromagnetic order
and is expected to be a good starting point to investigate the
resistivity anomalies~\cite{Dea98}. This double-exchange model, however, is
nothing else than the Kondo lattice model with ferromagnetic (positive)
exchange constant in the strong coupling limit.
In the CMR materials, the localized
$S=\frac{3}{2}$-spin of the model represents the more or less localized
manganese 
$3d$-$t_{2g}$ electrons, whereas the conduction band is formed by the $e_g$
electrons. The interband-exchange interaction is nothing else but the
intra-shell Hund's rule coupling. Since the $3d$-$e_g$ electrons of the
manganese form a relatively narrow band (theoretical results from
band-structure calculations: $1-2eV$~\cite{SPV96,PS96,SP98} and experimental
estimates: $3-4eV$~\cite{PCCBMCI96,SSKMFSTT97}) and Hund's coupling is
assumed to be
large, the model has to be taken in the intermediate to strong coupling
regime. There are few estimates about the value of the interaction
constant in the literature, e.g.\ $J\approx 1eV$~\cite{SPV96,OKIUAT95},
but these
are challenged as to be too small~\cite{MMS96}. Most theoretical
papers of the last years concerned with colossal magnetoresistance
assume classical spins
$S\rightarrow\infty$~\cite{Fur98pre,MLS95,MMS96,HV00,CMS00pre}. This
has been
justified by the assumption of $J
S\rightarrow\infty$~\cite{MMS96}. Although it is
true that the important energy scale is $J S$, there are much more
implications of $S\rightarrow\infty$ that are not justified in the
strong-coupling limit for a $S=\frac{3}{2}$ system. In several papers,
it was stated that ``\dots the $e_g$ electrons are oriented parallel to
the $t_{2g}$ spins.''~\cite{HV00} or equivalently ``\dots so one only
need consider configurations with $e_g$ electrons parallel to core
spins.''~\cite{MMS96}. We will show below using exact results as well as
several well-defined
approximation methods, that for $S=\frac{3}{2}$ there is a considerable
amount of spin-$\downarrow$
spectral weight located in the main region of the spin-$\uparrow$
states even for large
interaction strengths. The assumption of a half-metallic
state~\cite{SPHS00pre}, made in
the two citations above can therefore
never be met in the KLM with quantum spins and is merely an effect of
the (unphysical) limit of ``classical'' spins. The recently discussed
half-metallic behaviour of the manganites~\cite{ZK00pre} must have a
different origin.

However, for the opposite sign of $J$, exactly the assumed effect
happens in the 
strong-coupling limit: the lowest-lying excitations in the conduction band
density of states will be purely spin-$\downarrow$. This
already implies that results for the Kondo lattice model with $J>0$ and
$J<0$ cannot simply be reverted into the respective other case. The
change of sign changes the whole physics of the system.
For $J < 0$ an antiparallel
(``antiferromagnetic'')
alignment of the conduction band spin and the localized spin
lowers the internal
energy. For a sufficient band filling, this tends to
a screening of the local moments by conduction electrons, well-known
from the Kondo effect that refers to a single magnetic impurity in a
conduction electron sea. 
From this, the name ``Kondo lattice model'' was originally derived for the $J<0$
case.

We will further show that already for comparatively
low interaction strengths the spin-exchange interaction alone leads to
an opening of a gap in the density of states. This extraordinary
correlation effect could
give hints to the explanation of a recently discovered pseudogap in the
managese oxides~\cite{SDMKTH99pre}.

To prove our claims already laid out so far, we will first review two
important non-trivial exactly solvable limiting cases of the Kondo
lattice model in Sec.~\ref{sec:theo}. The first is the
\textit{zero-bandwidth limit} (``atomic limit'') 
where the bandwidth of the conduction band is set to
$W=0$~\cite{NM84}. The second
exactly solvable limiting case is the so-called
\textit{ferromagnetically saturated semiconductor}
~\cite{MM68,IM71,Ric71,SM81,AE82,AIK84,AI85,NDM85,ND85,NMR96,NRM97}. 
This is essentially
the zero-temperature limit of the model with vanishing electron
density and fully aligned spin system. In this limit, striking
correlation effects can be
observed and discussed. These limiting cases will already give clear
evidence to our propositions made above. In Sec.~\ref{sec:dmft} we will
present a new \textit{dynamical mean-field theory} (DMFT)-based approach
for the KLM with
$S=\frac{1}{2}$ spins.
To circumstantiate our theory, we 
will introduce three
more approximation schemes which also keep the 
spin as a quantum variable, not relying on the
classical spin limit. 
The first will be a \textit{second-order perturbation theory} (SOPT)
for the KLM based on  the projector operator formalism~\cite{Mor65,Mor66}.
The \textit{self-consistent CPA} (coherent potential
approximation) is a straightforward extension to the well-known CPA for
the KLM~\cite{Kub74,Nol79,NolBd7,EGK99,GE99b}. It starts from the
zero-bandwidth limit discussed before. The third approximation method
is a moment-conserving decoupling procedure for the equation of motion
of the single-electron Green function. This approximation scheme
continuously evolves from the
exactly solvable limit of the ferromagnetically saturated
semiconductor. It will be called 
\textit{Moment-Conserving Decoupling Approximation} (MCDA). 
The comparison of the results obtained by
these three methods and our DMFT scheme, each of
which starts from a different limit allows to 
evaluate the range of applicability of the new approach, and to
select the most
trustworthy common features of all methods to gain a reliable picture of
the physics of the ferromagnetic KLM.

\section{The Kondo lattice model and its many-body problem}
\label{sec:theo}

\subsection{Hamiltonian}
\label{sec:basic}
The Kondo-lattice model (or $s$-$f$ model) traces back the characteristic
features of the underlying physical system to an interband exchange
coupling of itinerant conduction electrons to (quasi-) localized
magnetic moments described by the following model
Hamiltonian~\cite{Nag74,Nol79}
\begin{eqnarray}
  \label{eq:hamiltonian}
  H=&H_s+H_{sf}+\left(H_U +H_{ff} \right)=\\
  =&\sum_{ij\sigma} T_{ij}
  c_{i\sigma}^{\dagger}c_{j\sigma}
  -J \sum_i \vec{\sigma}_i \cdot \vec{S}_i
+\left( \frac{1}{2}U\sum_{i,\sigma} n_{i\sigma} n_{i-\sigma} 
-\sum_{i,j} \hat{J}_{ij} \vec{S}_i \cdot \vec{S}_j
\right )\nonumber
\end{eqnarray}
$c^{\dagger}_{i\sigma}$ ($=\frac{1}{\sqrt{N}}\sum_{\vec{k}}
c_{\vec{k}\sigma}^{\dagger} e^{-i\vec{k}\cdot\vec{R}_i}$) and $c_{i\sigma}$
are, respectively, creation and annihilation
operators of a band electron being specified by the lower
indices. The hopping integrals $T_{ij}$ are connected by Fourier
transformation to the single-electron Bloch energies
$T_{ij}=\frac{1}{N}\sum_{\vec{k}} \epsilon(\vec{k}) e^{i\vec{k}
  \cdot(\vec{R}_i-\vec{R}_j)}$. 

The interband ($sf$) exchange with coupling strength $J$ is taken as an
intra-atomic interaction between
the conduction electron spin $\vec{\sigma}_i$ and the localized magnetic
moment represented by the spin operator $\vec{S}_i$.
For practical reasons it is sometimes convenient to use the second
quantization representation of the band electron spin $\vec{\sigma}_i$
which leads to the following form of the interband-interaction term:
\begin{equation}
  \label{eq:hamil-sf_2}
  H_{sf}=-\frac{1}{2} J \sum_{i,\sigma} \left ( z_{\sigma} S^z_i
    n_{i\sigma} + S^{\sigma}_i c^{\dagger}_{i-\sigma}c_{i\sigma}\right)
\end{equation}
Here we have used the abbreviations
$n_{i\sigma}=c_{i\sigma}^{\dagger}c_{i\sigma}$,
$z_{\uparrow(\downarrow)}=+1 (-1)$ and
$S_i^{\sigma}=S_i^x+iz_{\sigma}S_i^y$.
The first term in (\ref{eq:hamil-sf_2}) describes an Ising-like
interaction between the
z-components of the spin-operators, while the second term incorporates
spin exchange processes between the localized and the itinerant system.

The last two terms are an extension to the original model:
$H_U$ leads to the 'correlated Kondo lattice model', it introduces correlations
between the conduction electrons in form of a 'Hubbard-type'
interaction. We will include this term in some parts of the discussion
below. The second term, $H_{ff}$ represents a direct spin-exchange
between localized moments on different lattice sites. Although the $sf$
exchange can lead to an effective
RKKY interaction between the spins for non-vanishing band occupation, a
``superexchange''could also be included. In case of an empty conduction
band, this term becomes essential as source of magnetic order.
$\hat{J}_{ij}$ are the superexchange integrals. We state once more that the
original Kondo-lattice model (or $s$-$f$ model) is defined by
$H=H_s+H_{sf}$, only.
The additional
terms in Eq.~(\ref{eq:hamiltonian}) are used as
soon as physical requirements do not allow to neglect them.

If we are mainly interested in the conduction electron properties then
the single-electron Green function
$G_{ij\sigma}(E)=\langle \langle
  c_{i\sigma};c_{j\sigma}^{\dagger}\rangle\rangle_E$
is of primary interest. Its equation of motion reads for the
correlated KLM
\begin{equation}
  \label{eq:eom}
\fl    \sum_m (E \delta_{im} -T_{im}) G_{mj\sigma}(E) = \hbar
    \delta_{ij}
    -\frac{1}{2} J \left (z_{\sigma}
      I_{ii,j\sigma}(E)+F_{ii,j\sigma}(E)\right ) + U
    \Gamma_{iii,j\sigma}(E)
\end{equation}
The two types of interaction terms in~(\ref{eq:hamil-sf_2}) let appear
the ``spinflip function'' $F_{im,j\sigma}(E)=\langle \langle S_i^{-\sigma}
  c_{m-\sigma};c_{j\sigma}^{\dagger}\rangle \rangle_E$ and the ``Ising
  function''
$I_{im,j\sigma}(E)=\langle \langle S_i^z
  c_{m\sigma};c_{j\sigma}^{\dagger}\rangle \rangle_E$,
while the ``Hubbard-function'' 
$\Gamma_{ilm,j\sigma}(E)=\langle \langle c_{i-\sigma}^{\dagger} c_{l-\sigma}
  c_{m\sigma};c_{j\sigma}^{\dagger}\rangle \rangle_E$
comes into play only when the ``Hubbard-interaction'' (last term in Eq.~(\ref{eq:hamiltonian})) is
switched on.

The three ``higher'' Green functions on the right-hand side of
Eq.~(\ref{eq:eom}) prevent a direct solution of the equation of
motion. A formal solution for the Fourier-transformed single-electron
Green function,
\begin{equation}
  \label{eq:gf}
  G_{\vec{k}\sigma}(E)=\langle \langle c_{\vec{k}\sigma};
  c_{\vec{k}\sigma}^{\dagger} \rangle \rangle_E=
  \frac{\hbar}
  {E-\epsilon(\vec{k})-\Sigma_{\vec{k}\sigma}(E)} 
\end{equation}
defines the electronic selfenergy $\Sigma_{\vec{k}\sigma}(E)$ by the
ansatz 
$\langle \langle \left[c_{\vec{k}\sigma};
    H-H_s\right]_-;c_{\vec{k}\sigma}^{\dagger} \rangle
  \rangle_E = \Sigma_{\vec{k}\sigma}(E) G_{\vec{k}\sigma}(E)$
For the general case $\Sigma_{\vec{k}\sigma}(E)$ cannot be determined
rigorously.

Before introducing some approaches to the not exactly solvable many-body
problem of the KLM let us discuss in the next sections two rather
illustrative limiting cases which can help to test the unavoidable
approximations.

\subsection{The zero-bandwidth limit}
\label{sec:atomic}
Let us assume that the arbitrarily filled conduction band is shrinked to
an $N$-fold degenerate level $T_0$:
$\epsilon(\vec{k})\rightarrow T_0 \; \forall \vec{k}$.
Nevertheless, we consider the $f$-spin system as collectively ordered for
$T < T_{\rm c}$ by any direct or indirect exchange interaction. 
In this case, 
the hierarchy of equations of motion decouples exactly and
can rigorously be solved~\cite{NM84}. The resulting energies and respective spectral weights are
\begin{eqnarray}
\fl  E_1 = T_0 - \frac{1}{2} JS
  &\alpha_{1\sigma}=
  \frac{1}{2S+1}  \{ S+1+z_{\sigma}\langle
  S^z\rangle +\gamma_{-\sigma}- z_{\sigma}
  \Delta_{-\sigma}- 
  (S+1)\langle n_{-\sigma}\rangle \} \nonumber\\
  \label{eq:atomic_energies}
\fl  E_2 = T_0 + \frac{1}{2} J(S+1)
  &\alpha_{2\sigma} = \frac{1}{2S+1} \left \{ S-z_{\sigma}\langle
    S^z\rangle -\gamma_{-\sigma}+
    z_{\sigma}\Delta_{-\sigma}-S\langle n_{-\sigma}\rangle\right
  \}\\ 
\fl  E_3 = T_0 +U - \frac{1}{2} J(S+1)\;\;
  &\alpha_{3\sigma} = \frac{1}{2S+1} \left \{ S\langle
    n_{-\sigma}\rangle-
    \gamma_{-\sigma}+z_{\sigma}\Delta_{-\sigma}\right \}\nonumber\\ 
  \label{e4}
\fl  E_4 = T_0 +U + \frac{1}{2} JS
  &    \alpha_{4\sigma} = \frac{1}{2S+1} \left \{ (S+1)\langle
    n_{-\sigma}\rangle+\gamma_{-\sigma}-z_{\sigma}\Delta_{-\sigma}\right \}\nonumber 
\end{eqnarray}
The ``Hubbard-U'' in $E_3$ and $E_4$ indicates that these excitations are
bound to
a double occupancy of the respective lattice site. $E_1$ and $E_2$ appear when
our ``test electron'' enters an empty site. If it orients its spin
parallel to the local $f$-spin then the energy $E_1$ is needed. In case of an
antiparallel spin orientation a triplet or a singlet state is
formed. The first requires the energy $E_1$, the second $E_2$. The latter is
therefore two-fold degenerate. 
The spectral weights are,
contrary to the energy levels, strongly dependent on the
magnetization state of the $f$ system and the band filling.
For a complete solution one needs the average occupation number $\langle
n_{-\sigma}\rangle$ and the
mixed correlation functions $\gamma_{\sigma}=\langle
  S_i^{\sigma}c_{i-\sigma}^{\dagger}c_{i\sigma}\rangle$ and
  $\Delta_{-\sigma}=\langle S_i^zn_{i\sigma}\rangle$.
The evaluation can selfconsistently be done by use of the spectral
theorem for the
Green functions $G_{ii\sigma}(E)$, $I_{ii,i\sigma}(E)$ and
$F_{ii,i\sigma}(E)$ (cf.\ Ref.~\cite{NM84}).
It is interesting to observe that in any case from the four poles only
three do appear. For less than half-filled bands and $J>0$ ($J<0$)
$\alpha_{3\sigma}$ ($\alpha_{4\sigma}$) vanishes, and for more than half-filled
$\alpha_{2\sigma}$ ($\alpha_{1\sigma}$) does. 
It should be mentioned that the spectral weights
$\alpha_{i\sigma}$
do not explicitly depend on the coupling constants $J$ and $U$. That means,
on the one side, that even for $U = 0$ the $s$-$f$ interaction produces a
splitting into four not coinciding quasiparticle levels. On the other
hand, there is a striking dependence of $\Delta_{-\sigma}$ and $\langle
n_{-\sigma}\rangle$ on the sign of the $s$-$f$
coupling. That transfers to the spectral weights giving them some indirect
dependence on $J/|J|$. For $J >0$
and $J < 0$
the order of the energy
levels is different resulting via
the spectral theorem in different correlation functions. Note that the
mentioned dependence on $J$ concerns only the sign of $J$. The spectral
weights are not influenced by the absolute value $|J|$.

\subsection{The ferromagnetically saturated semiconductor}
\label{sec:magpol}
There is another very instructive limiting case that can be treated
rigorously. It concerns the situation of a single electron in an otherwise
empty conduction band at $T = 0$, when the local moment system is
ferromagnetically saturated. In this case the Coulomb
interaction is meaningless, the ``Hubbard-function''
$\Gamma_{ilm,j\sigma}(E)$ is identical 
to zero. In the zero-bandwidth limit, discussed in the last section, for
the $\uparrow$-spectrum all spectral weights disappear except for
$\alpha_{1\uparrow}=1$, while
for the $\downarrow$-spectrum the levels $E_1$ and $E_2$ survive with
$\alpha_{1\downarrow}=\frac{1}{2S+1}$ and
$\alpha_{2\downarrow}=\frac{2S}{2S+1}$.

For finite bandwidth the mentioned special case is that of a
ferromagnetically saturated semiconductor
~\cite{MM68,IM71,Ric71,SM81,AE82,AIK84,AI85,NDM85,ND85,NMR96,NRM97}.
In this case, the spin-$\uparrow$
quasiparticle density of states $\rho_{\uparrow}(E)$ is only rigidly
shifted compared to the ``free'' Bloch density of states.
\begin{equation}
  \label{eq:ro_up_magpol}
  \rho_{\uparrow}(E) \stackrel{T=0;n=0}{\longrightarrow} \rho_0(E+\frac{1}{2}JS)
\end{equation}
Consequently, the quasiparticle dispersion is undeformed
with respect to the Bloch energies,
$E_{\uparrow}(\vec{k})\rightarrow \epsilon(\vec{k})-\frac{1}{2}JS$.
\begin{figure}
  \begin{center}
    \includegraphics[width=0.6\textwidth]{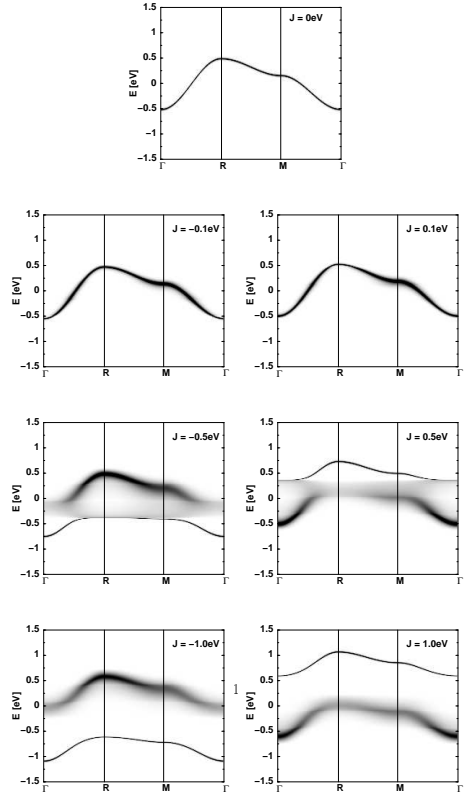}
  \end{center}
  \caption{Spin-$\downarrow$ spectral densities (quasiparticle
    bandstructures) of the ferromagnetically
    saturated semiconductor
    for $S=\frac{1}{2}$ and $J$ as
    indicated along several symmetry directions. The left (right)
    column displays the $J<0$ ($J>0$) case, the top-most picture shows
    the interaction-free spectral density.}
  \label{fig:mag_pol_spec}
\end{figure}
The $\downarrow$-spectrum is more complicated because a
$\downarrow$-electron has several possibilities to exchange its spin
with the antiparallel $f$ spins. Therefore, the
$\sigma=\downarrow$-spinflip function is
not at all trivial. However, its equation of motion decouples
exactly,
producing a closed system of equations which can be solved after Fourier
transformation for the single-electron Green function.
The corresponding selfenergy $\Sigma_{\vec{k}\downarrow}(E)$ reads:
\begin{eqnarray}
  \Sigma_{\vec{k}\downarrow}(E)&=\frac{1}{2} JS \left( 1+ 
    \frac{J B_{\vec{k}}(E)}{1-\frac{1}{2}JB_{\vec{k}}(E)} \right)\\
  B_{\vec{k}}(E) &= \frac{\hbar}{N}\sum_{\vec{q}} \left
    (E-\epsilon(\vec{k}-\vec{q})+\frac{1}{2}JS -\hbar
    \omega(\vec{q})\right)^{-1} 
\end{eqnarray}
$\hbar \omega(\vec{q})$ are the spin wave energies following from the
Heisenberg exchange $H_{ff}$ (cf.\ Eq.~(\ref{eq:hamiltonian})),
$\hbar\omega(\vec{q}) = 2S \left( \hat{J}(\vec{q}=0) - \hat{J}(\vec{q})\right)$. 
$\hat{J}(\vec{q})$ is the Fourier transform of the exchange integral
$\hat{J}_{ij}$. Usually the 
spin wave energies will be smaller by about two
orders of magnitude
than other typical energies of the system as the exchange constant $J$ or
the Bloch bandwidth $W$. As a general result the spectral density
$S_{\vec{k}\downarrow}(E)$
consists of two
structures corresponding to special elementary excitation processes of
the $\downarrow$ electron. There is a rather broad structure built up by
``scattering states'' which result from magnon emission by the original
$\downarrow$ electron. Thereby the excited electron reverses its spin
becoming a
$\uparrow$ electron. Such a process is possible only if there are
$\uparrow$ band
states within reach for the original $\downarrow$ electron to land after the
spinflip. The scattering states therefore occupy the same energy region
as the $\uparrow$-DOS (\ref{eq:ro_up_magpol}).

There is another possibility for the $\downarrow$ electron to flip its
spin. It can also be done by a repeated emission and reabsorption of a
magnon by the conduction electron resulting in a ``dressed'' particle
propagating through the lattice accompanied by a virtual cloud of
magnons. For not too small positive (negative) $J$ the energy of this ``dressed'' particle
lies above (below) the scattering spectrum giving even rise to a bound state,
i.e.\ to a quasiparticle with infinite lifetime which we call the
``magnetic polaron''. Outside the scattering region the polaron peak
manifests itself as a $\delta$-function. As soon as the peak dips into
the scattering part the polaron gets a finite lifetime after which it
decays into a $\uparrow$ electron plus
magnon. Figure~\ref{fig:mag_pol_spec} shows the down-spin quasiparticle
bandstructure as derived from the respective spectral density as a
density plot. The degree of blackening is a measure of the spectral
density magnitude. Sharp dark lines refer to pronounced peaks in the
spectral density representing quasiparticles with long life-time. For
weak coupling $|J S|< 0.1$ scattering processes smear out a little bit
the ``free'' dispersion but do not lead to strong deformations. However,
already for moderate couplings $|JS|\gtrsim 0.2$ one recognizes for some
$\vec{k}$ vectors the appearance of a sharp polaron dispersion. For
$J>0$ (right column) the magnetic polaron is stable on the high-energy
side of the $\uparrow$ spectrum, for $J<0$ on the low-energy side. In
addition the scattering spectrum is clearly visible taking away a great
part of the total spectral weight. In the antiferromagnetic KLM the
magnetic polaron represents the ground state configuration~\cite{TSU97}.
For still
rather moderate couplings of $|JS|\gtrsim 0.3$ the polaron dispersion
has split off over the full Brillouin zone. The magnetic polaron has an
infinite lifetime. It is surprising that even the broad scattering
structure is obviously bunched together as if it were a rather stable
quasiparticle. It is noteworthy to repeat that the results of
Fig.~\ref{fig:mag_pol_spec} are exact and free of any uncontrollable
approximation. So we have to expect these quasiparticle effects in real
systems, too.
\begin{figure}
  \begin{center}
    \resizebox{0.6\textwidth}{!}{
      \includegraphics{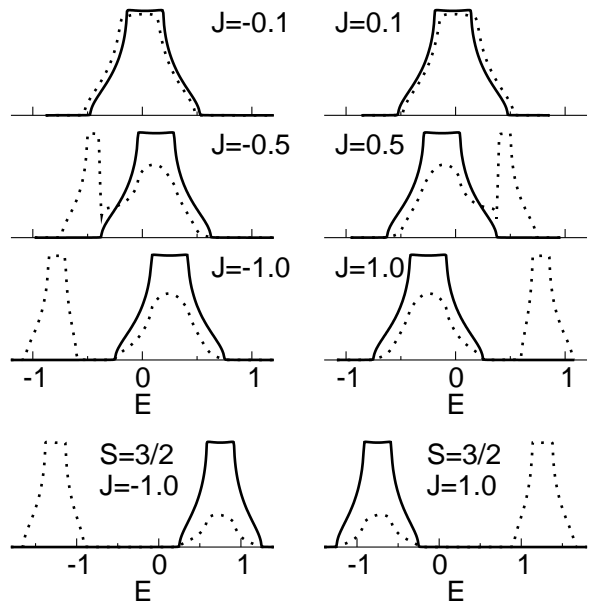}
      }
  \end{center}
  \caption{Spin-$\uparrow$ ($\downarrow$) density of states of the
    ferromagnetically saturated semiconducter as function of energy as
    solid (dotted) line.
    The upper three figures of each column show the DOS for a
    $S=\frac{1}{2}$ system with interaction strengths $|J|\in
    \{0.1;0.5;1.0\}$ corresponding to the spectral densities of
    Fig.~\ref{fig:mag_pol_spec}. The respective lowest figure shows a
    $S=\frac{3}{2}$ 
    system with $|J|=1.0$. The left (right) column corresponds to $J<0$
    ($J>0$).}
  \label{fig:mag_pol_dos}
\end{figure}
This holds also for the quasiparticle density of states (QDOS) plotted
in Fig.~\ref{fig:mag_pol_dos} for several exchange couplings
$|J|$. According to Eq.~(\ref{eq:ro_up_magpol}) the $\uparrow$-QDOS is
only rigidly shifted to higher (lower) energies for
$J<0$ ($J>0$). Correlation effects appear exclusively in the $\downarrow$
spectrum. For $|JS| \gtrsim 0.25$ a band splitting sets in. One of the
subbands occupies the same energy region as $\rho_{\uparrow}(E)$ being
therefore built up by the mentioned scattering states. In the
ferromagnetic ($J>0$) KLM it is the low energy part of the spectrum
containing a considerable amount of $\downarrow$-spectral weight. This
is not a specialty of the $S=\frac{1}{2}$ case or of weak and moderate
couplings exhibited in Fig.~\ref{fig:mag_pol_dos} but holds
equivalently, e.g., for $S=\frac{3}{2}$ and in the strong coupling
region ($|JS|\gtrsim1.0$, see last row in Fig.~\ref{fig:mag_pol_dos}). 
After the band splitting has set in the weights of the two
spin-$\downarrow$ subbands are close to the zero-bandwidth values
$\frac{1}{2S+1}$ for the lower, and $\frac{2S}{2S+1}$ for the upper part
($J>0$), i.e.\ independent of $J$.
The very often used
assumption, when the KLM is applied to the manganites, that the $e_g$
electron orients its spin parallel to the $t_{2g}$-$S=\frac{3}{2}$
spin~\cite{MMS96,HV00}, appears with respect to the exact results in
Fig.~\ref{fig:mag_pol_dos} rather questionable. In the antiferromagnetic
($J<0$) KLM the low energy quasiparticle subband consists of stable
polaron states which build the high energy part in the ferromagnetic
($J>0$) KLM. Although the $J>0$-QDOS and the $J<0$-QDOS appear as
mirror-images the implied physics turns out to be rather different.

\section{Dynamical mean-field theory}
\label{sec:dmft}
\subsection{Mapping onto an impurity model}
\label{sec:map}
It is not possible to directly apply the
methods of the dynamical mean-field theory~\cite{GKKR96} to the KLM. One
exception is the case of the classical-spin limit
($S\rightarrow\infty$)~\cite{Fur98pre,MMS96,HV00}, thus removing
the quantum nature of the spins. The effect of
quantum mechanics is strongest for $S=\frac{1}{2}$, but also for
$S=\frac{7}{2}$, they dramatically influence the
spectrum~\cite{NMR96,NRM97}. The assumption of classical spins for
$S=\frac{3}{2}$, as done in Refs.~\cite{MMS96,HV00}, needs
therefore careful analysis of the neglected effects.

Another possibility to derive a dynamical mean-field theory for the
Kondo lattice model is the fermionization of the localized spin
operators as suggested in~\cite{MO95}. This approach is, however limited
to $S=\frac{1}{2}$ systems (see also~\cite{PSK95,JPCL96,SBI99}). 
The first part of our
approach will closely follow Ref.~\cite{MO95}, but as discussed
below, will then differ from the cited reference.

The localized spins in the Hamiltonian~(\ref{eq:hamiltonian}) can be
expressed in terms of auxiliary fermion operators $f_{i\sigma}$
($f_{i\sigma}^{\dagger}$)~\cite{MO95}:
\begin{equation}
  \label{eq:fermionization}
  \vec{S}_i\rightarrow \vec{S}_i^{(f)}=\sum_{\sigma,\sigma'}
  f_{i\sigma}^{\dagger} 
  \vec{\tau}_{\sigma\sigma'} f_{i\sigma'}
\end{equation}
Here, $\vec{\tau}_{\sigma\sigma'}$ represents the Pauli matrices.
This is the same transformation  that led
to~(\ref{eq:hamil-sf_2}) where it was applied to the conduction
electron spin $\vec{\sigma}_i$.
The ``fermionization'' according to Eq.~(\ref{eq:fermionization})
implies the introduction of the constraint:
\begin{equation}
  \label{eq:constraint}
   Q_i=\sum_{\sigma}  n_{i\sigma}^{(f)}=
   \sum_{\sigma}f_{i\sigma}^{\dagger}f_{i\sigma}=1 
   \quad \forall i 
\end{equation}
The inclusion of this constraint via a Lagrange formalism corresponds to
the addition of a Hubbard-like interaction term for the
$f$-fermions to the Hamiltonian with the interaction
constant $U^{(f)}\rightarrow\infty$~\cite{MO95}. The one-particle energy of the
$f$-fermions is located at $-\frac{U^{(f)}}{2}$. The ``fermionized KLM''
takes the form:
\begin{equation}
  \label{eq:KLM+U}
\fl    H^{{\rm (ferm.)}}=\sum_{\vec{k},\sigma} \epsilon(\vec{k})
    c_{\vec{k}\sigma}^{\dagger}   c_{\vec{k}\sigma}
    - \frac{J}{2} \sum_i  \vec{\sigma}_i \cdot \vec{S}_i^{(f)} -
    \frac{U^{(f)}}{2}\sum_{i\sigma}f_{i\sigma}^{\dagger}f_{i\sigma}
    + \frac{U^{(f)}}{2}  \sum_{i\sigma} 
    n_{i\sigma}^{(f)}n_{i-\sigma}^{(f)}
\end{equation}
This Hamiltonian describes a system of two different kinds of
$S=\frac{1}{2}$ fermions coupled by a local spin-exchange
interaction. The $f$-fermions are additionally correlated via the
Hubbard-type of interaction to prevent double-occupancy.  This
Hamiltonian resembles the periodic
Anderson model (PAM)~\cite{hewson} in some way. In fact the only
difference in the operator structure is the ``coupling'' between
conduction band and $f$ state. In the PAM, this is simply a
kinetic-energy term (``hybridization''), whereas here, the two
subsystems are coupled by the spin exchange. 
Essentially, the ``fermionized KLM'' is a
rudimentary version of the general multi-band Hubbard model with (local)
inter-band interaction. For this model, the DMFT is discussed
by~\cite{GKKR96}.
It is straightforward to apply the standard methods of the
DMFT to map model~(\ref{eq:KLM+U}) onto an appropriate
impurity model with the corresponding self-consistency condition to
determine the parameters of the impurity model.
This leads to the following Hamiltonian for the impurity model
(single-site Kondo model, SSKM) with fermionized $f$-spins:
\begin{equation}
  \label{eq:SSKM-1}
\fl  H^{{\rm (ferm.)}}_{{\rm SSKM}}=\sum_{\vec{k},\sigma} \eta(\vec{k})
  c_{\vec{k}\sigma}^{\dagger}   c_{\vec{k}\sigma}
  - \frac{J}{2} \vec{\sigma}_0 \cdot\vec{S}^{(f)} -
  \frac{U^{(f)}}{2}\sum_{\sigma}f_{\sigma}^{\dagger}f_{\sigma}
  + \frac{U^{(f)}}{2}  \sum_{\sigma} 
  n_{\sigma}^{(f)}n_{-\sigma}^{(f)}
\end{equation}
with only one ``impurity'' site for the $f$-fermions. The spin-exchange
interaction acts only at this single site denoted by the site-index
$0$. It is advisable
to express the conduction band part of Hamiltonian~(\ref{eq:SSKM-1}) in
local space and single out the operators referring to the impurity site
$0$:
\begin{eqnarray}
  \label{eq:transformek}
  \sum_{\vec{k},\sigma} \eta(\vec{k})
  c_{\vec{k}\sigma}^{\dagger}c_{\vec{k}\sigma}=
  \sum_{i,j,\sigma} T_{ij} c_{i\sigma}^{\dagger}c_{j\sigma} =\\
  =\sum_{\sigma}T_{00} c_{0\sigma}^{\dagger}c_{0\sigma}+
  \sum_{{i\neq 0}\atop\sigma}T_{0i} \left(c_{0\sigma}^{\dagger}
    c_{i\sigma}+c_{i\sigma}^{\dagger}c_{0\sigma}\right) + 
  \sum_{{i,j\neq 0}\atop\sigma}T_{ij}
  c_{i\sigma}^{\dagger}c_{j\sigma}\nonumber 
\end{eqnarray}
For better readability, we will denote the construction operators at the
site $0$ with the symbol $d_{\sigma}$ ($d_{\sigma}^{\dagger}$), the
on-site energy will be denoted $T_{00}\rightarrow e_d$. Finally we
introduce a unitary transformation which diagonalizes the last term of
the second line in Eq.~(\ref{eq:transformek}). The transformed
fermion operators will be denoted $a_{k\sigma}$
($a_{k\sigma}^{\dagger}$), the hopping $T_{0i}\rightarrow V_{kd}$ and
finally $T_{ij}\rightarrow \tilde{\eta}(k)$ for $i\neq j$.
In the context of the DMFT, one does not need to know the explicit
structure of this unitary transformation since $\tilde{\eta}(k)$ and
$V_{kd}$ need not be known explicitly either. The parameters of the
single-site Kondo model will be determined by the self-consistency condition of the DMFT
(see below).

A direct solution of model~(\ref{eq:SSKM-1}) is, to our knowledge not
possible. However, it can be further simplified: The Hubbard term
originating from the constraint~(\ref{eq:constraint}) can be eliminated
by simply reversing the ``fermionization'' procedure which led from
Hamiltonian~(\ref{eq:hamiltonian}) to (\ref{eq:KLM+U}). The auxiliary fermion
operators get replaced by a local spin-$\frac{1}{2}$ operator at the
impurity site, and the constraint~(\ref{eq:constraint}) can be dropped.
This leads to the final version of the single-site Kondo model:
\begin{equation}
  \label{eq:SSKM}
\fl    H_{{\rm SSKM}}=\sum_{k,\sigma}\tilde{\eta}(k)
    a_{k\sigma}^{\dagger}a_{k\sigma} + 
    e_d\sum_{\sigma} d_{\sigma}^{\dagger}d_{\sigma}+ 
    \sum_{k,\sigma}
    V_{kd} \left( d_{\sigma}^{\dagger}a_{k\sigma} + a_{k\sigma}^{\dagger}
      d_{\sigma} \right) - \frac{J}{2} \vec{\sigma}^{(d)}\cdot \vec{S}
\end{equation}
This last step, which could be called ``de-fermionization'',
distinguishes our approach from the one used in Ref.~\cite{MO95} and
others. This ``de-fermionization'' ensures the exact fulfillment of the
constraint~(\ref{eq:constraint}), which could for example
only be kept on average ($\sum_\sigma \langle n^{(f)}_\sigma\rangle =1$
instead of $\sum_\sigma n^{(f)}_\sigma =1$) in~\cite{MO95}.

The parameters determining the conduction band in the
Hamiltonian~(\ref{eq:SSKM}),
namely $\tilde{\eta}(k)$ and $V_{kd}$ have to be specified according to
the DMFT self-consistency condition: The local conduction band Green
function (cf.\ (\ref{eq:gf})) should be equivalent to the $d$-operator Green
function of the single-site model, $\langle\langle
d_{\sigma};d_{\sigma}^{\dagger}\rangle\rangle =
G_{\sigma}^{(\rm d)}(E)$: 
\begin{equation}
  \label{eq:seco}
  \frac{1}{N}\sum_{\vec{k}} G_{\vec{k}\sigma}(E) =
  G_{\sigma}^{{\rm (d)}}(E) =
  \frac{\hbar}{E-e_d-\Delta_{\sigma}(E)-\Sigma_{\sigma}^{(d)}(E)}
\end{equation}
where the right equation follows from the formal
solution of the equation of motion of $G_{\sigma}^{(d)}(E)$. 
So instead of the
usual definition for the hybridization function, 
$\Delta(E)=\sum_{k}\frac{V_{kd}^2}{E-\tilde{\eta}(k)}$, it has to be 
determined so that Eq.~(\ref{eq:seco}) holds.
One will see below that the knowledge of $\Delta_{\sigma}(E)$, which can
become spin-dependent through this procedure, is sufficient to solve the
single-site Kondo model~(\ref{eq:SSKM}). Its dispersion
$\tilde{\eta}(k)$ and the hybridization parameter
$V_{kd}$ need not to be determined explicitly.

The DMFT, i.e.\ the mapping of the KLM onto the single-site model is,
except for the limit of infinite spatial dimensions, an approximation,
equivalent to the local approximation. In the exactly solvable case of
the ferromagnetic semiconductor (cf.\ Sec.~\ref{sec:magpol}) this is
equivalent to neglecting the magnon energies which
are assumed to be at least one order of magnitude smaller than
the energy scales under consideration here, e.g.\ bandwidth or $J$.

Next, we will introduce an approximative method to solve
the single-site Kondo model defined by Hamiltonian~(\ref{eq:SSKM}) for an arbitrary
hybridization function $\Delta_{\sigma}(E)$.

\subsection{Hybridization approximation}
\label{sec:hybrid}
In the following section, we derive an equation of motion-based method
to solve the
single-site Kondo model~(\ref{eq:SSKM}). Since in this section we deal only with this
model, we suppress all 
subscripts distinguishing between quantities in the lattice and the
single-site model. 

Starting point is the equation of motion for the $d$-Green function:
\begin{equation}
  \label{eq:eomgd1}
\fl  E G_{\sigma}^{(d)}(E)=\hbar+e_d G_{\sigma}^{(d)}(E) + \sum_k V_{kd}
  \langle\!\langle a_{k\sigma};d_{\sigma}^{\dagger}\rangle\!\rangle-
  \frac{J}{2} \sum_{\sigma} \left( z_{\sigma}
    I_{\sigma}(E)+F_{\sigma}(E) \right)
\end{equation}
where the higher Green functions $I_{\sigma}(E)=\langle\langle
d_{\sigma}S^z;d_{\sigma}^{\dagger}\rangle \rangle$ and $F_{\sigma}(E)=
\langle\langle d_{-\sigma}S^{-\sigma};d_{\sigma}^{\dagger}\rangle
\rangle$, corresponding to the Green functions $I_{im,j\sigma}(E)$
and $F_{im,j\sigma}(E)$ for the lattice case, are introduced. The
``mixed'' Green function $\langle\langle
a_{k\sigma};d_{\sigma}^{\dagger}\rangle\rangle$ can be eliminated by
investigating its equation of motion:
\begin{equation}
  \label{eq:mixedgd}
    \sum_k V_{kd} \langle\langle
    a_{k\sigma};d_{\sigma}^{\dagger}\rangle\rangle = 
  \sum_{k} \frac{V_{kd}^2}{E-\tilde{\eta}(k)} G_{\sigma}^{(d)}(E)
  =  \Delta(E)  G_{\sigma}^{(d)}(E)
\end{equation}
thereby defining the hybridization function $\Delta(E)=\sum_k
\frac{V_{kd}^2}{E-\tilde{\eta}(k)}$. This yields the final equation of
motion:
\begin{equation}
  \label{eq:eomgd}
  \left(E-(e_d+\Delta(E))\right) G_{\sigma}^{(d)}(E)=\hbar-
  \frac{J}{2} \sum_{\sigma} \left( z_{\sigma}
    I_{\sigma}(E)+F_{\sigma}(E) \right)
\end{equation}
Eq.~(\ref{eq:eomgd}) looks, except for the $\Delta(E)$ in the prefactor
of $G_{\sigma}^{(d)}(E)$, like the equation of motion
of the zero-bandwidth limit. The hybridization
function $\Delta(E)$ is due to the $V_{kd}$-term in the
Hamiltonian~(\ref{eq:SSKM}). This term prohibits an exact solution. In
fact, the
``hybridization'' via $V_{kd}$ is nothing else than the inter-site
hopping which was neglected in the zero-bandwidth limit. This term will
force us to make certain approximations in the determination of the
higher Green functions on the right-hand side of Eq.~(\ref{eq:eomgd}).
We will exemplify this using
the $I_{\sigma}(E)$-function. Its equation of motion reads
\begin{eqnarray}
  \label{eq:eomising}
  \fl (E-e_d)I_{\sigma}(E)=\hbar \langle S^z\rangle+\sum_k V_{kd}
  \langle\!\langle a_{k\sigma}S^z;d_{\sigma}^{\dagger}  
  \rangle\!\rangle-\\
  -\frac{J}{2}
  \left(z_{\sigma} \langle\!\langle 
  d_{\sigma}S^zS^z;d_{\sigma}^{\dagger}\rangle\!\rangle+
  \langle\!\langle
  d_{-\sigma}S^{-\sigma}S^z;d_{\sigma}^{\dagger}\rangle\!\rangle-
  z_{\sigma} \langle\!\langle
  d_{\sigma}d_{\sigma}^{\dagger}d_{-\sigma}
  S^{-\sigma};d_{\sigma}^{\dagger}\rangle\!\rangle\right) \nonumber
\end{eqnarray}
where on the one hand, higher ``impurity-site'' Green functions are
introduced, but on the other hand also a higher ``mixed'' Green
function, $\langle\!\langle
a_{k\sigma}S^z;d_{\sigma}^{\dagger}\rangle\!\rangle$. In analogy to the
one-particle mixed Green function, Eq.~(\ref{eq:mixedgd}),
we use the following substitution:
\begin{equation}
  \label{eq:hybridapprox}
  \sum_k V_{kd} \langle\langle
  a_{k\sigma} S^z ;d_{\sigma}^{\dagger}\rangle\rangle \rightarrow
  \Delta(E) \langle\langle
  d_{\sigma} S^z ;d_{\sigma}^{\dagger}\rangle\rangle =\Delta(E)
  I_{\sigma}(E) 
\end{equation}
The justification of this procedure can be found in analogy to the
``self-energy substitution''  known from other approximation methods for
the KLM (see e.g. ~\cite{NRM97}) by inspecting the
spectral representations of the relevant Green functions. This reveals
that all of them have the same single-particle poles, they differ only
in the respective weights given by matrix elements of the type 
$\langle E_n | A |E_m \rangle$. Here $|E_n\rangle$ are the energy
eigenstates and A represents one of the relevant operators:
$d_{\sigma}$, $a_{k\sigma}$, $d_{\sigma} S^z$ or $a_{k\sigma} S^z$. From
Eq.~(\ref{eq:mixedgd}) follows that the hybridization function
accounts for the differences of the matrix elements of the
first two operators ($d_{\sigma}$ and $a_{k\sigma}$). Assuming that
these differences are almost equal to
those of the matrix elements built up by the latter two operators leads
to the \textit{hybridization approximation}~(\ref{eq:hybridapprox}).

This is the only approximation necessary to decouple the hierarchy of
equations of motion. In the case of $S=\frac{1}{2}$ spins, there are
only 6 different
``impurity-site'' Green functions, whose equations of motion form a
closed set of equations. Besides the already introduced Green
functions $G_{\sigma}^{(d)}(E)$, $I_{\sigma}(E)$ and $F_{\sigma}(E)$,
these are
\begin{eqnarray}
\label{eq:gf_alle}
  F_{\sigma}^{(1)}(E)& =\langle\!\langle d_{-\sigma}
  d_{-\sigma}^{\dagger} d_{\sigma} S^{z};d_{\sigma}^{\dagger}
  \rangle\!\rangle \nonumber \\
 F_{\sigma}^{(2)}(E)& =\langle\!\langle d_{-\sigma}
  d_{\sigma}^{\dagger} d_{\sigma} S^{-\sigma};d_{\sigma}^{\dagger}
  \rangle\!\rangle \\
  D_{\sigma}(E)&= \langle\!\langle
  d_{-\sigma}d_{-\sigma}^{\dagger}d_{\sigma}
  ;d_{\sigma}^{\dagger}\rangle\!\rangle \nonumber
\end{eqnarray}
Several expectation values, introduced into the theory via
the inhomogeneities of the equations of motion can be expressed in
terms of the above-mentioned Green
functions, a self-consistent solution has to be found.

At this point, let us comment on the reliability of this
approximation. Although only one approximation enters our decoupling
procedure (cf.\ Eq. (\ref{eq:hybridapprox})), it still has to be seen as
an uncontrollable approximation meaning that there is no true small
parameter. There are two non-trivial limiting cases where the
replacement~(\ref{eq:hybridapprox}) becomes exact:
The first is the limit $V_{kd}\rightarrow0$, representing the zero-bandwidth
KLM of Sec.~\ref{sec:atomic}. But already a small, but
finite bandwidth in the KLM could lead to any (unknown) expression for
$\Delta(E)$ implying that we can not necessarily assume $V_{kd}$ to be
small any more. The second limit is the ``classical spin'' limit where
the ``spin variable'' $S$ has no operator properties any more. Here, the
replacement of Eq.~(\ref{eq:hybridapprox}) reduces
to~(\ref{eq:mixedgd}). However since we are interested in the general
case with finite $S$ and bandwidth, we have to confirm the
trustworthiness of this approximation
by a comparison with other well-tested methods.

\section{Other approximation methods}
\label{sec:approx}
We are now going to introduce three further approximation methods for
the KLM. These are known from literature, their strengths and
weaknesses have been identified.
The three methods
differ substantially
with respect to the theoretical assumptions made for an approximate
solution of the KLM. 
Common features following from these procedures and the
above-introduced DMFT scheme
should then give some credit of reliability ,
in particular when they additionally fit the exact limiting cases
discussed in Sec.~\ref{sec:theo}.

\subsection{Second-order perturbation theory}
\label{sec:sopt}
An application of the usual diagrammatic perturbation theory to the
Kondo model cannot be performed due to the absence of Wick's theorem.
Only for low temperatures (spin-wave approximation)~\cite{WW70} or in the
classical-spin limit~\cite{RHB67}, this method is applicable.
The  projection operator formalism of Mori~\cite{Mor65,Mor66} is
better suited for the Kondo model.
It has been used successfully to describe correlation
effects in the Hubbard model in the weak coupling regime~\cite{BJ90}.
The general formula for the second-order contribution
$\Sigma^{(2)}_{\vec{k}\sigma}(E)$ 
can be found there 
(Eq. (3.12) in Ref.~\cite{BJ90}).
To allow for a better comparison with the other approximations in this
paper, we further approximate the self-energy taking only $\vec{k}$
averaged occupation numbers into account (local approximation).

\subsection{Self-consistent CPA}
\label{sec:self-consistent-cpa}
Next, we want to introduce a modification of the well-known 
``coherent potential approximation'' (CPA)~\cite{VKE68} for the KLM.
The CPA is a standard many-body approach 
that starts from a fictitious alloy in analogy to the interacting
particle system. Starting point may be the zero bandwidth limit of
Sec.~\ref{sec:atomic}. We think of a four-component alloy each
constituent of which
is characterized by one of the energy levels $E_i$ in
Eq.~(\ref{eq:atomic_energies}). The
spectral weights $\alpha_{i\sigma}$~(\ref{eq:atomic_energies}) are then
to be interpreted as the
``concentrations'' of the alloy components as seen by a propagating
$\sigma$-electron. The fictitious alloy for a $\sigma$-electron is built
up by the
local moments and by the frozen ($-\sigma$) electrons. The CPA-selfenergy of
the $\sigma$-electron is found by the well-known formula~\cite{VKE68}.
\begin{equation}
  \label{eq:cpa}
    0=\sum_{p=1}^{4}
    \alpha_{p\sigma}\frac{E_{p}-\Sigma_{\sigma}(E)-T_0}
  {1-G_{ii\sigma}(E) (E_{p}-\Sigma_{\sigma}(E)-T_0)}
\end{equation}
As a consequence of the single-site aspect of the CPA the resulting
selfenergy is wave-vector independent.
According to Eq.~(\ref{eq:atomic_energies}) the ``concentrations''
$\alpha_{i\sigma}$
depend on a sum of the ``higher'' correlation
functions $I_{ii,i\sigma}(E)$ and $F_{ii,i\sigma}(E)$,
which can rigorously be expressed
by the single-electron Green function. 
\begin{equation}
  \label{eq:spectheo_delta}
    \gamma_{\sigma} + z_{\sigma} \Delta_{\sigma} = \frac{-1}{\hbar\pi}
    \frac{1}{N}
    \sum_{\vec{k}} \int_{-\infty}^{+\infty} dE f_-(E)
    \left(E-\epsilon(\vec{k})\right) \Im G_{\vec{k}\sigma}(E)
\end{equation}
The shortcomings of the CPA-procedure
lie on hand. The one is the same as that in the conventional alloy
analogy of the Hubbard model, namely the assumption of frozen ($-\sigma$)
electrons.
This is partially removed by our proposed modification of the standard
CPA procedure for the KLM, namely the selfconsistent calculation of the
higher correlation functions $\gamma_\sigma$ and $\Delta_\sigma$ via
Eq.~(\ref{eq:spectheo_delta}) as well as the band-occupation 
$\langle n_{-\sigma}\rangle$ via the spectral theorem for the
one-electron Green function.
Maybe even more serious
in the case of the KLM is the blocking of repeated spin exchange with
the local moment system. Magnon emission or absorption is not
involved. So we cannot expect that the CPA-treatment correctly
reproduces the exact limiting case of Sec.~\ref{sec:magpol}.
However,
some general
information about the quasiparticle bandstructure might be possible, in
particular in the strong coupling (``split band'') regime. By construction
the method yields the correct zero-bandwidth limit.

\subsection{The moment-conserving decoupling approach (MCDA)}
\label{sec:moment-cons-eom}
A Green function method which takes the spin dynamics correctly into
account has been proposed in Ref.~\cite{NRM97}. 
For details about this approach, we refer the read to the cited paper,
here we summarize the result shortly

This decoupling approach yields finally a selfenergy of the following
structure: 
\begin{equation}
  \label{eq:sestruc}
  \Sigma_{\vec{k}\sigma}(E)= -\frac{1}{2}J z_{\sigma} \langle S^z\rangle
  + \frac{1}{4} J^2 D_{\vec{k}\sigma}(E)
\end{equation}
The first term is linear in the coupling $J$ and proportional to the $4f$
magnetization $\langle S^z\rangle$. It just represents the result of a
mean-field
approximation being correct in the weak-coupling limit. The second term
in~(\ref{eq:sestruc}) contains all the spin exchange processes which may
happen. It is
a complicated functional of the selfenergy itself. So~(\ref{eq:sestruc})
is not at all
an analytical solution but an implicit equation for the selfenergy. We
do not present here the lengthy expression for $D_{\vec{k}\sigma}(E)$
referring the reader
for further details to Ref.~\cite{NRM97}. It should be mentioned, however, that
$D_{\vec{k}\sigma}(E)$ contains several expectation values which must be
fixed to get a
self-consistent solution. No problems arise with the mixed correlation
functions $\gamma_{\sigma}=\langle
  S_i^{\sigma}c_{i-\sigma}^{\dagger}c_{i\sigma}\rangle$ and $\Delta_{\sigma}=\langle S_i^zn_{i\sigma}\rangle$.
They can
rigorously be expressed by the spin flip function $F_{ii,i\sigma}(E)$
and the Ising function $I_{ii,i\sigma}(E)$ defined previously.
Both functions are already involved in the procedure, so that no further
approximations are necessary to fix $\gamma_{\sigma}$ and
$\Delta_{\sigma}$. For pure local-moment
correlations such as $\langle S_i^z\rangle$, $\langle
S_i^{\pm}S_i^{\mp}\rangle$,\dots, however, a special treatment is
necessary, e.g. as described in Ref.~\cite{NRM97}.

\section{Results}
\label{sec:results}
In the following section, we discuss the results obtained for the DMFT
and the three
approximation schemes of Sec.~\ref{sec:approx} for the ferromagnetic
Kondo lattice model with $S=\frac{1}{2}$.
Since we are interested in the reaction of the conduction band
due to the magnetic order of the spin system, we have not calculated
the latter self-consistently. Instead, we have simulated the magnetic
order by determining $\langle S^z \rangle$ using a Brillouin
function. 
Temperatures are given in units of $T_{\rm c}$.  Within
the CPA, our choice of $\langle S^z \rangle$ can lead to unphysical
results. Namely in the case of $\langle S^z \rangle \rightarrow S$, some
of the weights~(\ref{eq:atomic_energies}) can become negative.
There is an
upper bound for $\langle S^z\rangle$~\cite{NM84}. We therefore had to
limit $\langle S^z\rangle$  to
$\langle S^z\rangle\lesssim 0.33$ for some of the CPA calculations.
Within the DMFT calculations, we
experienced severe numerical problems which forced us to introduce a
further approximation: ``mean-field''-decoupling the $F_{\sigma}^{(1)}(E)$ and
$F_{\sigma}^{(2)}(E)$ functions
simplifies the system of equations of motion:
\begin{eqnarray}
\fl    F_{\sigma}^{(1)}(E) \approx& (1-\langle n_{-\sigma}\rangle)
    \Gamma_{\sigma}(E) + \langle S^z \rangle D_{\sigma}(E)+
    + \langle
    (1-n_{-\sigma}) S^z \rangle G_{\sigma}^{(d)}(E)\nonumber \\
  \label{eq:unrestHF}
 \fl  F_{\sigma}^{(2)} (E) \approx& \langle n_{\sigma}\rangle F_{\sigma}(E)
  - \langle  S^{-\sigma}d_{\sigma}^{\dagger}d_{-\sigma}\rangle
  G_{\sigma}^{(d)}(E) 
\end{eqnarray}
All DMFT-results presented below were obtained using the
hybridization approximation in combination with this
unrestricted-mean-field decoupling of $F_{\sigma}^{(1)}(E)$ and
$F_{\sigma}^{(2)}(E)$.

\begin{figure}
  \begin{center}
      \includegraphics[width=0.7\textwidth]{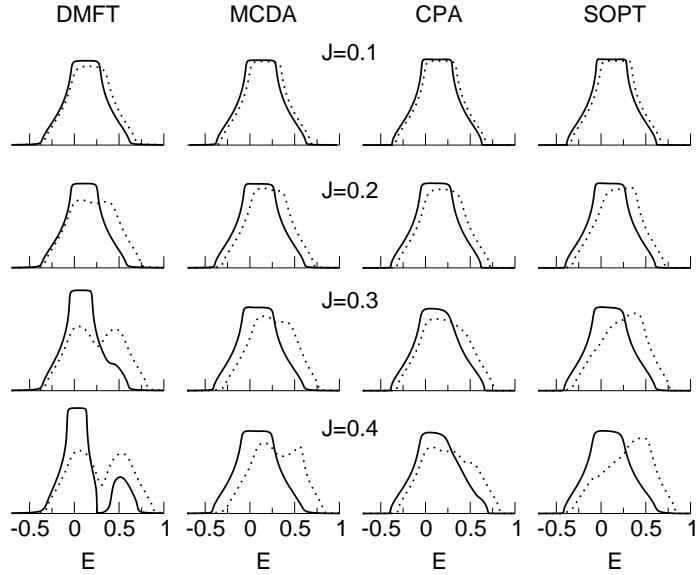}
  \end{center}
  \caption{Conduction band density of states (DOS) obtained via the DMFT,
    MCDA, CPA and SOPT (from left to right) for different values of
    $J$ as function of energy. The temperature is set
    to $T=0$, therefore $\langle S^z \rangle =0.5$  (CPA: $\langle
    S^z\rangle =0.33$, see text) and the electron
    density $n=0.5$, the chemical potential is at $\mu=0$;
    Solid line: spin-$\uparrow$, dotted line spin-$\downarrow$ DOS}
  \label{fig:sf_t0}
\end{figure}
In all calculations, the conduction band is described by a
tight-binding DOS for a simple-cubic lattice structure~\cite{Jel69}
of unit width ($W=1eV$). The Curie temperature is
taken as $T_{\rm c}=250K$.

\begin{figure}
  \begin{center}
      \includegraphics[width=0.7\textwidth]{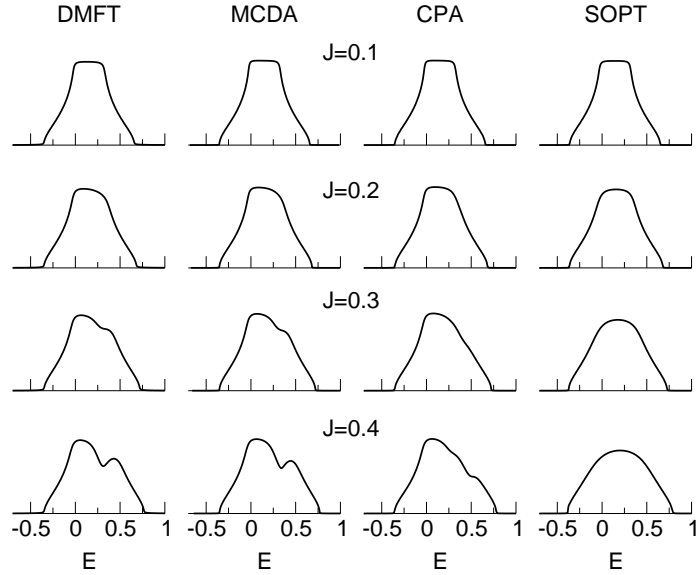}
  \end{center}
  \caption{Same as in Fig.~\ref{fig:sf_t0}, but for $T=T_{\rm c}$,
    therefore $\langle S^z \rangle =0$.}
  \label{fig:sf_tc}
\end{figure}

The quasiparticle densities of
states (DOS) for quarter-filling and different values of $J$ are plotted
in Figs.~\ref{fig:sf_t0}
and~\ref{fig:sf_tc} for $T=0$ and $T=T_{\rm c}$, respectively. As
indicated, the columns correspond to DMFT, MCDA, CPA and SOPT,
respectively (from left to right). 

We will begin the discussion with the DMFT results shown in the left
column of Fig.~\ref{fig:sf_t0}. A small value of $J$ leads to a
spin-dependent shift of the spin-$\uparrow$ and $\downarrow$ DOS, as one
would obtain by a simple mean-field decoupling,
i.e.\ by replacing $S^z$ by its mean value $\langle S^z\rangle$ in
Hamiltonian~(\ref{eq:hamiltonian}). With increasing $J$, the DOS
show some striking correlation effects: first a
broadening, later the onset of a splitting of the band can be observed.
Whereas in general, the correlation effects are stronger for spin
$\downarrow$ (indicated by a stronger quasiparticle damping), the
splitting is, in contrast to the other two methods, more pronounced in
the spin-$\uparrow$ DOS. However, here
the split-of (upper) peak has much less spectral weight than the
original peak which, except for a band-narrowing still resembles
strongly the free conduction band DOS.

The picture is very similar in the MCDA. Again, a
mean-field like
spin-dependent bandshift is observed for small $J$. On increasing $J$, the
spin-$\downarrow$ DOS broadens, and a two-peak structure emerges. The
spin-$\uparrow$ DOS remains, except for a small tail at its upper edge,
unchanged. This behavior can easily be understood by comparing with the
special case of the ferromagnetically saturated semiconductor
as discussion in Sec.~\ref{sec:magpol} since the MCDA develops
continously into
this special case for $n\rightarrow 0$ and $T\rightarrow 0$. The
spin-$\downarrow$ DOS splits
into a scattering part (low energies) and the polaron-like part at
higher energies above the conduction band. Unlike the $n=0$ case,
however, there are some, but weak modification in the $\uparrow$ DOS,
namely the 
above mentioned tail at its upper edge. This can be interpreted as a
scattering contribution for spin-$\uparrow$ electrons. The origin of
this is the finite number of spin-$\downarrow$ electrons.

Also the CPA DOS show a remarkably similar picture as the previously
discussed theories. Again, there is the
mean-field shift for small $J$. For larger values
of $J$, the splitting of the spin-$\downarrow$ DOS sets in similarly to the
other two methods. Here, this can be contributed to the two
single-occupancy quasiparticle energies $E_1$ and $E_2$ known from the
zero-bandwidth limit
(cf.\ Sec.~\ref{sec:atomic}).
For $n \neq 0$ a third quasiparticle energy carries
non-vanishing weight,
namely the $E_4$ peak. This is located between the two other peaks
since we switched off the conduction band Coulomb interaction $U$
(cf.\ Eq.~(\ref{eq:hamiltonian})).
For
$J=0.4$, the appearance of this third peak in between the two main peaks is
vaguely visible.
A remarkable difference to the DMFT and MCDA results is the fact, that
the spin-$\uparrow$ and scattering part of the spin-$\downarrow$ DOS do
not cover the same energy range. This
shortcoming of the CPA is due to the neglection of spin-dynamics
(magnons) as mentioned in Sec.~\ref{sec:self-consistent-cpa} (cf.\
Ref.~\cite{AI85}).

The SOPT results are, for the plotted values of $J$ in the weak- and
intermediate coupling regime, not too far away from the other
results. For small $J$, where the SOPT becomes by definition reliable,
the mean-field shift as in the other methods is clearly
observable. Similar to the other methods, the deformation of the
spin-$\downarrow$ DOS is much stronger than that of the spin-$\uparrow$
DOS. However, for larger $J$, the SOPT never shows a true band
splitting, its range of validity is certainly restricted.

For $T=T_{\rm c}$, where spin-symmetry is re-established, the DMFT, MCDA
and CPA give a similar overall picture.
The resulting DOS shows, already for $J\gtrsim 0.3$, a two-peak
structure (DMFT and
MCDA), in the CPA a third peak is dimly noticeable.
The most remarkable observation is the near-coincidence of
the DMFT and MCDA results for all $J$ values. The transition from the
(clearly distinct) $T=0$ DOS to the (resembling) $T=T_{\rm c}$ DOS is
continuous for both theories.
The SOPT, however, has to be seen as a complete
failure for a paramagnetic system already for relatively small 
$J\approx 0.3$. This can already be read of the formula for calculating
the self-energy, which is more
or less trivial. The corresponding results cannot be connected to any of
the exactly solvable 
limiting cases (cf.\ Sec.~\ref{sec:sopt}).

A sound-standing interpretation of the observations is not difficult
since the MCDA 
can be traced back to the exactly solved limiting case of 
the ferromagnetically saturated semiconductor (cf.\
Sec.~\ref{sec:magpol}) and the CPA to that of
the zero-bandwidth limit discussed in Sec.~\ref{sec:atomic}.
The SOPT, of course, becomes reliable for small $J$.
Although all four presented methods are of approximate nature and their
results do, at least for intermediate-to-large values of $J$, differ,
some common
properties emerge:
For small $J$, the behavior is genuine mean-field like, a bandshift
proportional to $\pm J \langle S^z \rangle$ is observed. For larger $J$,
the onset of  a band splitting occurs. At $T=0$ this primarily affects
the spin-$\downarrow$
DOS. This fact is understandable by examination of the
ferromagnetically saturated semiconductor
where the band splitting was discussed in terms of a scattering band and
the magnetic polaron. In that case the spin-$\uparrow$ DOS remains,
except for a
simple shift, unaffected by
the interaction simply because spin-flip of spin-$\uparrow$ electrons is
suppressed in this case. Now for finite $n$, this is only approximately
true. The spin-$\downarrow$ DOS still shows strong correlation effects,
but also the spin-$\uparrow$ DOS is affected due to the finite number of
spin-$\downarrow$ electrons in the system. 
This manifests itself differently in the various methods: in the MCDA, a
tail is seen at the upper edge of the spin-$\uparrow$ DOS, in the CPA, a
shoulder develops, and in the DMFT approach, a band splitting is observable.
From this we conclude
that correlation effects are
much more pronounced in the DMFT than in the other two theories.
At $T=T_{\rm c}$, the dip indicating the onset of the band splitting is
still existing for all but the SOPT method. In CPA, MCDA and the DMFT
results this splitting is of similar size, which can easily be read off
the CPA where it is simply given by the difference of the respective
energies from the zero-bandwidth limit (see
Eq.~(\ref{eq:atomic_energies})).
The two most
pronounced peaks correspond to the single-occupancy quasiparticle
energies $E_1$ and $E_2$, the band splitting is therefore approximately
$\Delta E= J (S+\frac{1}{2})$.

This is in contrast to the results obtained by
dynamical mean-field theory for the $S\rightarrow\infty$ KLM (KLM with
classical spins). The
emerging picture for classical spins is the following~\cite{Fur98pre}: At
$T=0$, the DOS is
characterized by
a mean-field like ``Zeeman'' splitting between the bands of both spin
directions. With increasing temperature, at each of the respective
spin-$\uparrow$ or $\downarrow$ band, spectral weight of the opposite
spin direction appears, until finally at $T=T_{\rm c}$, the system
becomes paramagnetic. The splitting into two subbands separated by
$\Delta E=JS$ stays constant.
Comparing our results with the $S\rightarrow\infty$ ones,
phenomenologically the DOS's show completely different characteristics at
$T=0$, especially in the spin-$\downarrow$ band. For $T=T_{\rm c}$, the
DOS's look quite similar. However, the physics behind the scenes turn out
to be completely different as can be seen, e.g., by the inconsistent size of
the band splitting $\Delta E$. Whereas in the classical-spin limit, the
splitting
is always due to a mean-field like ``on-site Zeeman splitting'', in the
case of quantum
spins, the various elementary excitations as discussed in the case of
the ferromagnetically saturated semiconductor, are responsible for the
sub-band structure.
For the ferromagnetically saturated system, the
neglection of spin-flip processes and magnons in the
$S\rightarrow\infty$ calculation lead to the picture of a
half-metal~\cite{Fur98pre}. This does not apply for any finite value of
$S$ and is therefore clearly an artifact of the $S\rightarrow\infty$
limit.

Let us shortly remark on the situation with anti-ferromagnetic coupling
($J<0$)~\cite{MeyerDiss}. All four approximation methods presented here
do not show any signs of Kondo screening. An investigation of the
special low-temperature physics of the model with $J<0$ is therefore not
possible. However, some remarks about the behavior of the model for
$T>T_{\rm K}$ can be made: In general the excitation spectra are broader
than in the $J>0$ case. The resulting exchange splitting of $\Delta
E\approx J(S+1)$ is also larger than for $J>0$ which can already be seen
in the zero-bandwidth limit. This is again in sharp contrast to the
$S=\infty$ results, where the size of the splitting is independent of
the sign of $J$.

\section{Conclusions}
\label{sec:conclu}
In this paper, we investigated the Kondo lattice model (KLM) focusing on
the model with
positive (ferromagnetic) exchange
constant $J$. We discussed two exactly solvable, but nevertheless
non-trivial limiting cases as well as four different approximation
methods.
The results obtained by these methods, maybe except for the perturbation
theory,
compare generally reasonably well. Sometimes even nearly perfect matches occur
(cf.\ Fig.~\ref{fig:sf_tc}). In general, the differences between the
methods are smaller for the paramagnetic than the ferromagnetic system.

From the comparison of common features of these approximation methods in
combination with the exact results, the following picture emerges for
the quasiparticle structure of the ferromagnetic Kondo lattice model:
For small $|J|$, a mean-field like shift of $\uparrow$ and
$\downarrow$ DOS is visible.
Already for intermediate coupling strengths, a band splitting
of size $\Delta E=J(S+\frac{1}{2})$ occurs. The relevant energy scale for
the splitting is $JS\approx 0.3$.
In the case of ferromagnetic saturation ($T=0$), the splitting is more
pronounced in
the $\downarrow$ than in the $\uparrow$ DOS. 
This band-splitting should not be confused with the splitting found in
the limit of ``classical spins'' ($S\rightarrow\infty$). There, the
splitting is simply due to a mean-field like Zeeman-splitting and
therefore of the size $\Delta E=JS$. Contrary
to that, our
results clearly show that the two emerging subbands can be traced back to
the two elementary excitations known from the limit of the
ferromagnetically saturated semiconductor (see above).
The magnetic saturation of the spin-system suppresses these processes for
spin-$\uparrow$ electrons which explains the stronger footprint of the
correlations in the spin-$\downarrow$ DOS.
For finite temperatures, the magnetic polaron generally remains a well
defined quasiparticle, represented by a rather sharp Lorentzian peak in
the spectral density. Now the spin-$\uparrow$ electrons can also
participate in spin-exchange processes since the localized spins are not
fully aligned any more. The spin-symmetric DOS at $T\gtrsim T_{\rm c}$
also show the characteristic splitting.
Our results also confirm the fundamental differences between the $J>0$
and the $J<0$ case of the Kondo lattice model:
It can be read of both
exactly solvable limiting cases that the ground state will be different
depending on the sign of $J$. The approximative approaches of
Secs.~\ref{sec:dmft} and~\ref{sec:approx} do not allow a deeper
investigation of the model with $J<0$ due to the inability to reproduce
the special low-temperature properties of that model (``Kondo physics'').
Another important conclusion from our calculations can be drawn: There
is always finite spin-$\downarrow$ 
spectral weight in the region of the spin-$\uparrow$ DOS.
This spectral weight does not disappear in the limit
$J\rightarrow\infty$ but only in the limit $S\rightarrow\infty$
(``classical spins'').
For the ferromagnetic KLM ($J>0$), this implies that the KLM for
$S=\frac{3}{2}$ and large $J$, corresponding to the situation found in
manganites,  will not
be a half-metal for $T=0$, contrary to the predictions of
$S\rightarrow\infty$ calculations.

\ack
Financial support by the \textit{Volkswagen-Foundation} within the
project \textit{``Phasendiagramm des Kondo-Gitter-Modells''} 
is gratefully
acknowledged. 
This work also benefitted from
the financial support of the \textit{Sonderforschungsbereich SFB 290}
(``Metallische d\"unne Schichten: Struktur, Magnetismus und elektronische 
Eigenschaften'') of the Deutsche Forschungsgemeinschaft.
One of us (D.\ M.\ ) wants to thank the
\textit{Friedrich-Naumann-Foundation} for supporting his work.


\end{document}